\documentstyle[12pt,epsfig]{article}
\begin{document}

\begin {center}
{\Large The mass of the $\sigma$ pole.}
\vskip 3mm

{D.V.~Bugg\footnote{email address: D.Bugg@rl.ac.uk}},   \\
{Queen Mary, University of London, London E1\,4NS, UK}
\end {center}
\vskip 2.5mm

\begin{abstract}
BES data on the $\sigma$ pole are refitted taking into account
new information on coupling of $\sigma$ to $KK$ and $\eta \eta$.
The fit also includes Cern-Munich data on $\pi \pi$ elastic
phases shifts and $K_{e4}$ data, and gives a pole position of
$500  \pm 30 - i(264 \pm 30)$ MeV.
There is a clear discrepancy with
the $\sigma$ pole position recently predicted by Caprini et al. using
the Roy equation.
This discrepancy may be explained naturally by
uncertainties arising from inelasticity in $KK$ and
$\eta \eta$ channels and mixing between $\sigma$ and
$f_0(980)$.
Adding freedom to accomodate these uncertainties gives an optimum
compromise with a pole position of $472 \pm 30 - i(271 \pm 30)$ MeV.

\vspace{5mm}

\end{abstract}

\section {Introduction}
BES data on $J/\Psi \to \omega \pi^+ \pi^-$ display a conspicuous
low mass $\pi \pi$ peak due to the $\sigma$
pole [1].
It was observed less clearly in earlier DM2 [2] and E791 data [3].
The BES data are reproduced in Fig. 1.
The band along the upper right-hand edge of the Dalitz plot, Fig. 1(a),
is due to the $\sigma$ pole. There is a clear peak in the $\pi \pi$
mass projection of Fig. 1(b) at $\sim 500 $ MeV; the fitted $\sigma$
contribution is shown by the full histogram of Fig. 1(d). Other
large contributions to the data arise from $f_2(1270)$ and
$b_1(1235)$, which appears in the $\omega \pi$ mass projection of Fig.
1(c).

\begin {figure}  [htp]
\begin {center}
\vskip 2cm
\epsfig {file=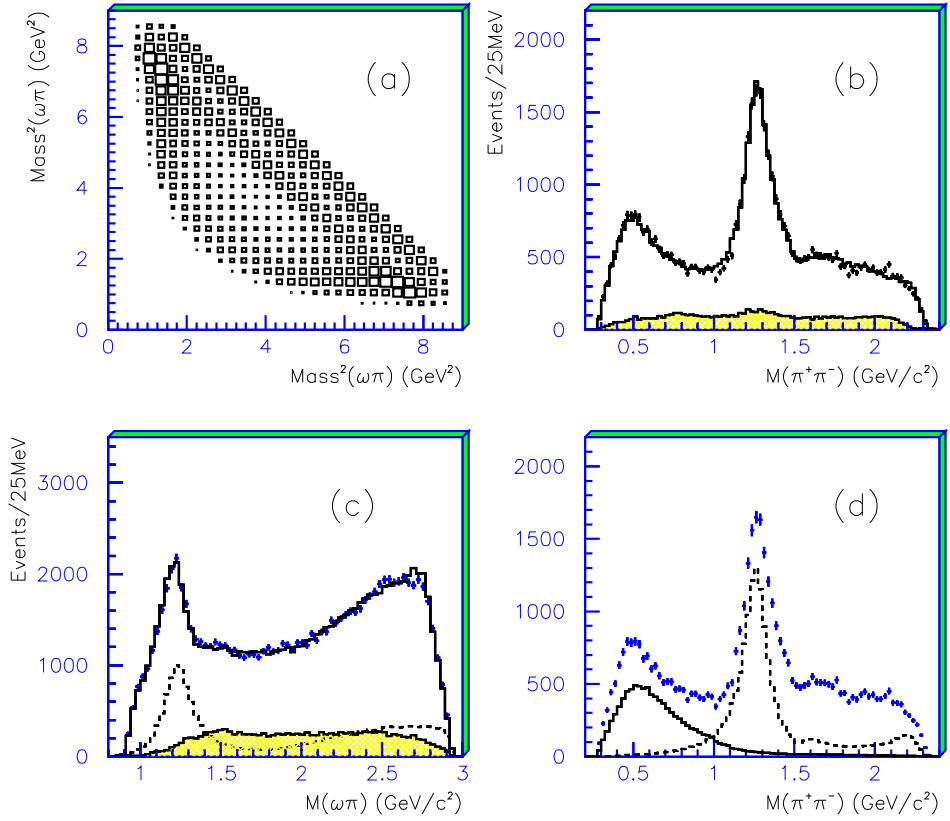,width=13cm}\
\vskip -6cm
\caption {BES data of Ref. [1]. (a) Dalitz plot;
(b) $\pi \pi$ mass projection: the upper histogram shows fit (ii)
of Table 1, the lower histogram shows experimental background;
(c) $\omega \pi$ mass projection; full histograms are as in (b) and
the dashed histogram shows the coherent sum of both $b_1(1235)\pi$
contributions; (d) $\pi \pi$ mass projections from $\sigma$ (full
curve) and spin 2 (dashed).}
\end {center}
\end {figure}

At the time of the BES analysis, little was known about the
coupling of $\sigma$ to $KK$ and $\eta \eta$, and these channels
were omitted from the fit to the data.
Since then, couplings to $KK$ and $\eta \eta$ have been determined by
fitting (a) all available data on $\pi \pi \to KK$ and $\eta \eta$,
(b) Kloe and Novosibirsk data on $\phi \to \gamma (\pi ^0 \pi ^0)$ [4].
All these data agree on a substantial coupling of $\sigma$ to $KK$.
The first objective of the present work is to refit the $J/\Psi \to
\omega \pi \pi$ data including this coupling, following
exactly the procedure of the BES publication.
The outcome is to move the $\sigma$ pole position from
$541 \pm 39 -i(252 \pm 41)$ MeV reported in the BES paper to
$500 \pm 30 - i(264 \pm 30)$ MeV.

Meanwhile, Caprini, Colangelo and Leutwyler (denoted hereafter
as CCL) have made a prediction of the $\sigma$ pole position [5].
Their calculation is based on the Roy equation for $\pi \pi$
elastic scattering [6], which embodies constraints of analyticity,
unitarity and crossing symmetry.
This approach has the merit of including driving forces from the
left-hand cut due to exchange of $\rho$, $f_2$ and $\sigma$.
They also apply tight constraints from Chiral Perturbation Theory
(ChPT) on the S-wave scattering lengths $a_0$ and $a_2$ for isospins
0 and 2.
Experiment alone determines only the upper side of the pole well;
this has led to speculation that fits without a pole might succeed
in fitting the data [7].
CCL's use of crossing symmetry and analyticity gives a precise
determination of the magnitude and phase of the S-wave amplitude on
the lower side of the pole and leaves no possible doubt about its
existence.
They clarify the fact that $\pi \pi$ dynamics are
fundamental to creating the pole.
They quote rather small errors: $M_\sigma = 441 ^{+16} _{-8} -
i(272 ^{+9}_{-12.5})$ MeV.
There is a rather large discrepancy between this prediction and
BES data.
A second objective of the present work is to trace the origin of
this discrepancy.
In outline, what emerges is as follows.

From CCL results for $T_{11}(s)$, the T-matrix for $\pi \pi$
elastic scattering, one can predict what should appear in
BES data.
The prediction will be shown below on Fig. 4 by the chain
curve.
From a glance at this figure, one sees a
significant disagreement with the experimental points, which are
deduced directly from BES data.
The questions which arise are as follows:
\begin {itemize}
\item {(i) are the hypotheses used in the BES analysis wrong or
questionable?}
\item {(ii) is something
missing from the calculation with the Roy equation?}
\item {(iii) can the calculation of CCL be fine-tuned in order to come
into agreement with the experimental data? }
\end {itemize}

The discrepancy lies in the mass range 550 to 950 MeV.
In this range, the analysis is sensitive to assumptions about
inelasticities due to $KK$ and $\eta \eta$ channels.
These are not known with sufficient accuracy at present
and allow freedom in the analytic continuation of couplings to
these channels below their thresholds.
The amplitude for $KK \to \pi \pi$ goes to zero at the $KK$ threshold;
it  also has an Adler zero at $s \simeq 0.5m_\pi^2$.
In between, it has a  peak near 500 MeV providing a natural
explanation of the additional peaking required by BES data.
This can readily explain the discrepancy, as shown by the full curve
on Fig. 4.
It may be regarded as a fine-tuning of the solution of the Roy equation.
It will be shown that changes to $\pi \pi$ phase shifts up to 750 MeV
are $<1.2^\circ$, well below experimental errors.
In summary, the Roy equation and Chiral Perturbation Theory provide
the best description of $\pi \pi$ scattering near threshold (and below),
while the BES data provide the best view of the upper side of the
$\sigma $ pole.

The layout of the paper is as follows.
Section 2 introduces the equations.
Section 3 discusses the prediction of CCL and subsection 3.1
explains what is missing from their Roy solution.
Section 4 describes fits to BES data.
Section 5 discusses possible alternative explanations of BES data
and Section 6 summarises conclusions.

\section {Equations}
   The $\pi \pi$ elastic amplitude may be written
\begin {equation}
T_{11}(s) = (\eta e^{2i\delta }-1)/(2i) = N(s)/D(s),
\end {equation}
where $N(s)$ is real and describes the left-hand cut;
$D(s)$ describes the right-hand cut.
The numerator contains an Adler zero at $s = s_A \simeq 0.41m^2_\pi$:
\begin {equation}
N(s) = (s - s_A)f(s),
\end {equation}
where $f(s)$ varies slowly with $s$.
The standard relation to the S-matrix is
$T_{11}=(S_{11}-1)/2i$;
$T_{11}$ has contributions from
$f_0(980)$, $f_0(1370)$ and $f_0(1500)$ as well as $\sigma$.
For elastic scattering, these contributions are combined by multiplying
their S-matrices, to satisfy unitarity; this implies adding phases
rather than amplitudes.

In the BES publication, it was shown that their data may be fitted
with the same $1/D(s)$ for the $\sigma$ component as $K_{e4}$ data [8]
and Cern-Munich data [9] within errors.
The phase of the $\pi \pi$
S-wave in BES data has the same variation with $s$ as the $\sigma$
component in elastic scattering within experimental errors of
$\sim 3.5^\circ$ from 450 to 950 MeV [10].
The BES data may be fitted by an amplitude $\Lambda /D(s)$; here,
$\Lambda$ is taken as a constant, since the left-hand cut of
$J/\Psi \to \omega \sigma$ is very distant.
Possible doubts about this assumption will be discussed later.
Channels $\pi \pi$, $KK$, $\eta \eta$ and $4\pi$ will be labelled
1 to 4.
The parametrisation of the $\sigma$ is given by
\begin {eqnarray}
T_{11}(s) &=& M\Gamma _1(s)/[M^2 - s - g^2_1
\frac {s-s_A}{M^2-s_A}z(s) - iM\Gamma _{tot}(s)] \\
M\Gamma_1(s) &=& g^2_1\frac {s-s_A}{M^2-s_A}\rho_1(s) \\
g^2_1(s) &=& M(b_1 + b_2s)\exp [-(s - M^2)/A] \\
j_1(s) &=& \frac {1}{\pi}\left[2 + \rho _1  ln_e \left(
\frac {1 - \rho _1}{1 + \rho _1}\right) \right] \\
z(s) &=& j_1(s) - j_1(M^2)  \\
M\Gamma_2(s) &=& 0.6g^2_1(s)(s/M^2)\exp (-\alpha |s-4m^2_K|)\rho_2(s) \\
M\Gamma_3(s) &=& 0.2g^2_1(s)(s/M^2)\exp (-\alpha |s-4m^2_\eta|)
\rho_3(s) \\
M\Gamma_4(s) &=& Mg_4\rho_{4\pi}(s)/\rho_{4\pi }(M^2) \\
\rho _{4\pi}(s) &=& 1.0/[1 + \exp (7.082 - 2.845s)].
\end {eqnarray}
D(s) is the denominator of eqn. (3).
Below the $KK$ threshold, $Im~D = -N(s)$ and $\tan \delta =
-Im~D/Re~D$, from which one can deduce $D(s)$.
Since $D(s)$ is analytic, it is subject to a normalisation uncertainty
which is constant within the errors of $T_{11}(s)$, in practice
2-3\%.

The function $j_1(s)$ is obtained from a dispersion integral over
the phase space factor $\rho _{\pi \pi }(s) = \sqrt {1 - 4m^2_\pi/s}$.
An  important point about $z(s)$ is that it is well behaved at $s = 0$
and eliminates the singularity in $\rho (s)$.
It makes the treatment of the $\pi \pi$ channel fully analytic - an
improvement on earlier work.
In the $4\pi$ channel, a dispersion integral is in principle required
but is small below 1 GeV and can be absorbed into the fit to $g_1^2(s)$.
In eqn. (11), $4\pi$ phase space is approximated empirically by a
combination of $\rho \rho$ and $\sigma \sigma$ phase space [11] (with
$s$ in GeV$^2$); $\Gamma (4\pi)$ is set
to zero for $s <16m^2_\pi$.
The effect of the $4\pi$ channel on the $\sigma $ pole is only 2 MeV
and is not an issue.

It is assumed that the Adler zero is a feature of the full
$\pi \pi$ amplitude.
The factor $(s - s_A)/(M^2 - s_A)$ of eqn. (4) introduces this Adler
zero explicitly.
Eqn. (5) is an empirical form used earlier in fitting BES data;
$b_1$, $b_2$ and $A$  are fitted constants.
The factors $s/M^2$ in $\Gamma _2$ and $\Gamma _3$ of eqns. (8) and (9)
approximate the Adler zeros closely at $s=0.5m^2_\pi$,
$0.5m^2_K$ and $0.5m^2_\eta$ and remove the square root singularity in
$\rho _2$ and $\rho_3$ of  eqns. (8) and (9).
The factors 0.6 and 0.2 for $g^2_{KK}$ and  $g^2_{\eta \eta }$ have
been fitted to data on $\pi \pi \to KK$ and $\eta \eta$ and on
$\phi \to \gamma (\pi ^0 \pi ^0)$ [4]. These fits also determine
$\alpha _{KK} = \alpha _{\eta \eta} = 1.3$ GeV$^2$ above the
thresholds.

A detail is that the factor $(s-s_A)/(M^2-s_A)$ is also used for
$\Gamma _1$, $\Gamma _2$ and $\Gamma _3$ of $f_0(980)$, $f_0(1370)$ and
$f_0(1500)$.
Otherwise, parameters of $f_0(980)$ are taken from
BES data on $J/\Psi \to \phi \pi ^+\pi ^-$ and $\phi K^+K^-$ [12].
The $f_0(1500)$ and $f_0(1370)$ are fitted with Flatt\'e formulae and
parameters given in Ref. [4].
The combined contribution of $f_0(980)$, $f_0(1370)$ and $f_0(1500)$
to the scattering length is 8\%.

An important question is how to parametrise the continuation of
$KK \to  \pi \pi$ and $\eta \eta \to \pi \pi$ amplitudes below their
thresholds, i.e. in K-matrix notation the elements $K_{12}$ and
$K_{13}$.
They can in principle be determined from dispersion relations
for each amplitude.
The factors $\rho _2$ and $\rho _3$ in eqns. (8) and (9) are kinematic
factors.
Below the thresholds, they are continued analytically as
$i\sqrt{4m^2_i - s}$.
If $\rho _2$ is factored out of the $KK \to \pi \pi$ amplitude, there
is a step at threshold in the imaginary part of the surviving
amplitude.
This step leads to a cusp in the real part of the amplitude at
threshold, i.e. a change of slope in eqns. (8) and (9).
An evaluation of the dispersion integral generates
a result for the real part of the amplitude below threshold
close to an exponential falling as $\exp [-\alpha (4m^2_i - s)]$.

\begin {figure}  [htb]
\begin {center}
\vskip -10mm
\epsfig {file=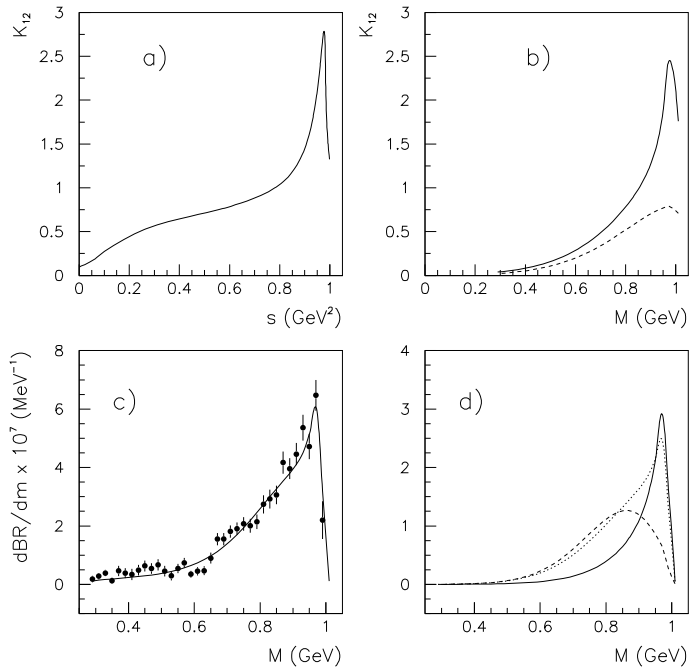,width=10cm}\
\vskip -4mm
\caption {(a) the magnitude of the $KK \to \pi \pi$ amplitude from
B\" uttiker et al. [13];
the result from my fit to Kloe data (full curve) and the $\sigma$
contribution alone (dashed); (c) fit to Kloe
$\phi \to \gamma (\pi ^0\pi ^0)$ data after background subtraction;
(d) contributions from $f_0(980)$ (full curve), $\sigma$ (dashed) and
interference (dotted).}
\end {center}
\end {figure}

A calculation of $K_{12}$ has been made along these lines by
B\" uttiker, Descotes-Genon and Moussallam using the Roy equations
for $\pi K \to \pi K$ and $\pi \pi \to KK$ [13].
Their result is reproduced in Fig. 2(a).
The peak is due to $f_0(980)$ and the low mass tail comes from the
analytic continuation of $\sigma \to KK$.
Using their $K_{12}$ in fitting BES data does lead to effects
with the right trend, but not with sufficient accuracy to make a good
prediction.
They remark that their calculation is uncertain because of
discrepancies between available sets of data on $\pi \pi \to KK$.
My estimate is that errors of Caprini et al from the
dispersion relation are at least a factor 2 too small.

My conclusion is that $K_{12}$ and $K_{13}$ presently need to be
fitted empirically.
There is however experimental information which helps decide an
appropriate parametrisation:
data on $\phi \to \gamma (\pi ^0 \pi ^0)$ from Kloe [14].
In Ref. [4], those data are fitted using the standard $KK$
loop model of Achasov and Ivanchenko [15].
Parameters of $f_0(980)$ are fixed to those determined by BES data
[12].
The fit requires a substantial additional amplitude for $KK \to \pi
\pi$ through the $\sigma$.
Empirically this contribution is well fitted with the exponential of
eqn. (8) with $\alpha = 2.1$ GeV$^{-2}$ below the $KK$ threshold
and an Adler zero at $s = 0.5m^2_K$.
[The same $\alpha$ is assumed for $\eta \eta \to \pi \pi$.]
Fig. 2(c) reproduces from Ref. [4] the fit to Kloe
data; Fig. 2(d) shows the $f_0(980)$ and $\sigma$ components and the
interference between them.
Fig. 2(b) compares my fit with that of B\" uttiker et al.;
the dashed curve shows the $\sigma$ contribution.

\section {The Roy solution of CCL}
CCL make a prediction of the $\pi \pi \to \pi \pi $ S-wave amplitude
and  deduce the $\sigma$ pole position from it.
The inputs to their calculation are [5,16,17]:
\newline i) the Roy equation, which accounts for both left and
right-hand cuts in $\pi \pi$ elastic scattering;
\newline ii) the precise lineshape of $\rho (770)$ from $e^+e^- \to \pi
^+\pi ^-$ data;
\newline iii) predicted value from ChPT for S-wave scattering lengths
$a_0 = (0.220 \pm 0.005)m_\pi ^{-1}$,
$a_2 = (-0.0444 \pm 0.0010) m_\pi ^{-1}$ [17];
\newline iv) the $\pi \pi$ phase shift at 800 MeV with an error of
$^{+10}_{-4}$ deg;
\newline v) my elasticity parameters $\eta$ above the $KK$ threshold;
\newline vi) minor dispersive corrections for masses above 1.15 GeV,
their matching point.
\newline Note that experimental phase shifts are not fitted except for
constraint (iv).

CCL have kindly supplied a tabulation of their $T_{11}(s)$.
My first step is to reproduce this solution using eqns. (1)--(11).
This is simply a fit to their fit, i.e. an explicit
algebraic parametrisation.

They do not explicitly separate $\sigma$ and $f_0(980)$.
This leads to uncertainty in how the $f_0(980)$ is being fitted.
In the vicinity of the $\sigma$ pole, an issue is the magnitude of
the low mass tail of $f_0(980)$, which affects what is left as the
remaining $\sigma$ amplitude.
CCL omit $f_0(1370)$ and $f_0(1500)$ contributions and
$\sigma \to 4\pi$.

The fit shown in column (i) of Table 1 is made in two steps.
Firstly a fit to their $f_0(980)$ is made from 800 to 1150
MeV.
It requires $M = 0.970$ GeV, $g^2_1 = 0.146$ GeV$^2$, somewhat
smaller than 0.165 GeV$^2$ from BES data on
$J/\Psi \to \phi f_0(980)$ [12].
This step also reveals that their $\sigma \to KK$ and $\eta \eta$
components fall from these thresholds at least as fast as $s^{2.25}$;
their contributions below 800 MeV are very small.
In the second step, the mass range below 800 MeV is refitted, in order
to minimise the sensitivity of $\sigma$ parameters
to the $f_0$ mass region.
Empirically, this second step can reproduce CCL phases only if
(a) $\sigma \to KK$ and $\eta \eta$ contributions are omitted and
(b) the $f_0 \to \pi \pi$ amplitude is multiplied by a factor
$(s/M^2)^n$, where $n$ optimises at 1.55.
Their phases are then reproduced everywhere up to 750 MeV within
errors of $0.12^\circ$, as shown by the full curve of Fig. 3.
This figure shows fitted values minus the phase shifts of
CCL for the full $\pi \pi$ elastic amplitude.
The pole position is 449-i269 MeV, 8 MeV higher in mass
than the CCL pole.

\begin{table}[htb]
\begin {center}
\begin{tabular}{cccc}
\hline
                    & (i)   & (ii)  & (iii) \\\hline
$M$ (GeV)           & 1.038 & 0.958 & 0.953 \\
$b_1$ (GeV)         & 1.082 & 1.201 & 1.302 \\
$b_2$ (GeV$^{-1})$  &-0.016 & 0.684 & 0.340 \\
$A$(GeV$^2)$        & 1.179 & 2.803 & 2.426  \\
$g_{4\pi }$ (GeV)   & 0     & 0.014 & 0.011 \\
Pole (GeV) & $0.449 - i0.271$ & $0.500-i0.264$ & $0.472-i0.271$ \\\hline
\end{tabular}
\caption{Parameters of fits discussed in the text.}
\end {center}
\end{table}

\begin {figure}  [htb]
\begin {center}
\vskip -17mm
\epsfig{file=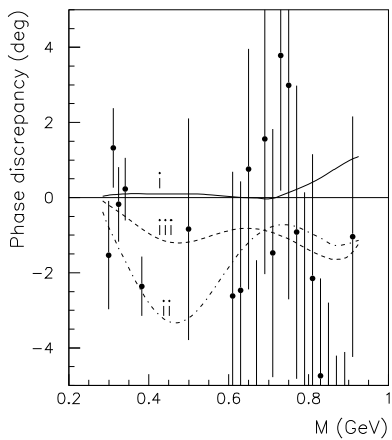,width=7cm}\
\vskip -5mm
\caption{Fits (i)--(iii) of Table 1 minus the
predicted phases of CCL. The full curve is fit (i) to
CCL phases, the dashed curve is fit (iii), the compromise fit between
CCL phases and BES data, the chain curve is fit (ii) to
$K_{e4}$ and Cern-Munich phases and BES data; points with errors
illustrate departures from CCL phases of $K_{e4}$, Cern-Munich data and
a point at the $K$ mass.}
\end {center}
\end {figure}

The essential conclusion is that their contributions from $f_0(980) \to
\pi \pi$, $KK \to \pi \pi$ and $\eta \eta \to \pi \pi$ are cut off very
sharply at high mass, and the whole of the $\pi \pi$ amplitude
below about 500 MeV is being fitted by the $\sigma$ alone.
This is hardly surprising.
It is well known that $f_0(980)$ is driven largely by forces in
the $KK$ channel, not by the left-hand cut of the $\pi \pi$
channel.
This question will be discussed more fully below.
For the moment, it is sufficient to remark that the pole immediately
moves up to 467 MeV if the $f_0(980) \to \pi \pi$ amplitude is
multiplied only by the factor $s/M^2$ of eqns. (8) and (9) and the
$KK \to \pi \pi$ and $\eta \to \pi \pi$ amplitudes of
the $\sigma$ are treated with the falling exponential fitted to Kloe
data.

\begin {figure}  [htb]
\begin {center}
\vskip -1cm
\epsfig{file=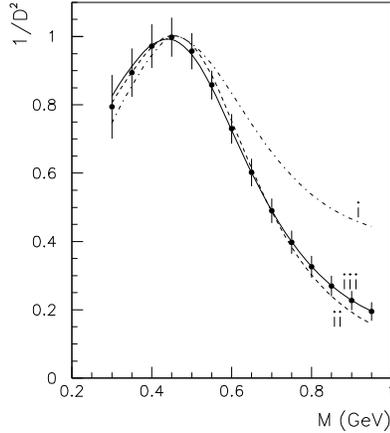,width=7cm}\
\vskip -8mm
\caption{$|1/D(s)|^2$ from BES data ; the chain curve is
fit (i) to $T_{11}$ values of CCL.
The dashed and full curves are from fits
(ii) and (iii) of Table 1.}
\end {center}
\end {figure}

    Fig. 4 displays values of $|1/D(s)|^2$ from
BES data as points with errors, normalised to 1 at their peak; they
are obtained by dividing out 3-body phase space from data of
Fig. 1(b).
Note that the normalisation is arbitrary -- it is the
$s$-dependence which matters.
There is a significant disagreement with the prediction from fit (i)
to CCL phase shifts (chain curve); this discrepancy is far
larger than the 2-3\% errors in deducing $D(s)$.
The essential point is that BES data fall more rapidly
from 600 to 950 MeV than CCL's result.
Fig. 5 shows the poor fit to BES data using their $D(s)$.
Note that the discrepancy is {\it not } a question
of the extrapolation of amplitudes to the pole.
There is a direct conflict for physical values of $s$ between CCL
and the fit to BES data.
However, bearing in mind that this is a theoretical prediction, the
closeness to data is still remarkable.

\begin {figure}  [htp]
\begin {center}
\vskip -12mm
\epsfig{file=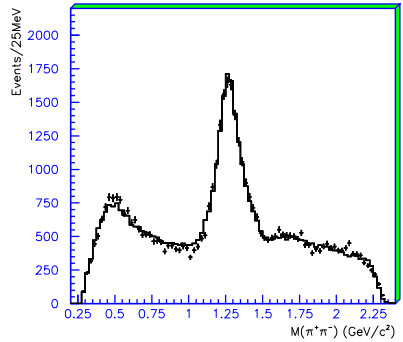,width=5.5cm}\
\vskip -4mm
\caption{Fit to BES data using $1/D(s)$ derived from CCL phases.}
\end {center}
\end {figure}

\subsection {Missing elements in the Roy solution}
The Roy equation is in principle exact.
How is it possible to modify the solution?
In Nature, the Roy equation `knows' about coupling of
$\pi \pi$ to $KK$ and $\eta \eta$, as well as the $f_0(980)$,
$f_0(1370)$ and $f_0(1500)$ resonances.
If $\pi \pi$ phase shifts were known with errors of a
small fraction of a degree, it would be possible to deduce
these resonances and their couplings.
However, that is not the present situation.
In reality, there are subtle features in the processes
$\pi \pi \to KK$ and $\eta \eta$ arising from meson
exchanges and also from mixing between $\sigma$ and $f_0$.

Although the $\sigma$ pole appears remote from the $KK$ threshold,
one must remember that the phase shift it produces reaches $90^\circ$
close to the position of $f_0(980)$, so multiple scattering is a
maximum there.
This includes terms of the form $\sigma \to \pi \pi $ (or $KK$)
$\to f_0(980)$ leading to mixing.
Anisovich, Anisovich and Sarantsev show [18] that this mixing obeys
the Breit-Rabi equation, so $\sigma$ and $f_0(980)$ behave as a pair of
coupled oscillators.

A recent paper with van Beveren, Rupp and Kleefeld [19] shows
that the nonet of $\sigma$, $\kappa$, $a_0(980)$ and $f_0(980)$
may be generated by coupling of mesons to a $q\bar q$ loop.
The $\sigma$ is generated by the $n\bar n$ loop and the
$f_0(980)$ by the $s\bar s$ loop.
Without mixing between $\sigma$ and $f_0$, one can only account
for $<10\%$ of the observed magnitude of $f_0(980) \to \pi \pi$.
Substantial mixing is required to produce the observed $\pi \pi$
width.
Using their programme, I have tried varying the mixing angle
and its $s$-dependence to examine perturbations of the
$\pi \pi$ amplitude.
Since the Schr\" odinger equation is solved, the amplitudes are fully
analytic.
In the mass range 700-1150 MeV, one sees subtle correlated changes
in $\pi \pi$ phases and the inelasticity parameter $\eta$.
These are not present in the calculation of CCL.
The Roy equation is simply a dispersion relation between real and
imaginary parts of the $\pi \pi \to \pi \pi$ amplitude.
Its solution for the real part is only as good as the input for
the imaginary part.
They derive results for $f_0(980)$, $KK \to \pi \pi$ and
$\eta \eta \to \pi \pi$ from my inelasticity
parameters of Ref. [4]; but those are based upon
simple Flatt\' e parametrisations of $f_0$ and $\sigma$, without any
dynamics due to mesons exchanges and mixing.

Calculations in Ref. [19] show that the $f_0$ pole is anchored to the
$KK$ threshold.
What happens as the mixing is varied is that an $f_0 \to \pi \pi$
amplitude reaching down to the $\pi \pi$ threshold is generated.
It is questionable exactly what its precise mass dependence will be.
The best conjecture one can make at present is that it will contain
the Adler zero of the full $\pi \pi$ amplitude, but otherwise
behaves like a normal resonance.
That is the conjecture adopted here.
Calculations with the model of van Beveren et al show that the
$\sigma$ pole can be affected by the mixing with $f_0(980)$ by
up to 50 MeV.
The reason for this is that the mixing alters the $s$-dependence
of the $\pi \pi$ amplitude.
The pole position is determined by the Cauchy-Riemann relations
as one moves off the real $s$-axis.
It lies far below the mass at which the $\pi \pi$ phase shift
reaches $90^\circ$.
Because of this long lever arm, even a small change of the
slope of $\pi \pi$ phases v. mass can move the pole a surprisingly
long way.
This is why including an $f_0(980)$ contribution making up $7\%$ of
the scattering length moves the pole position up to 467 MeV.

This degree of freedom is missing from the calculation of CCL.
My conclusion is that it is legitimate to introduce into the
fit small systematic perturbations in eqns. (4), (8) and (9) arising
from inelasticities in $KK$ and $\eta \eta$ channels and also mixing
between $\sigma$ and $f_0$.
This secures agreement with BES data with only small changes in
$\pi \pi$ phase shifts.

\section {Refit of BES data}
In the next fit, BES data are refitted together with
$K_{e4}$ and Cern-Munich data as in the BES
publication, but using the new equations of Section 2.
The small $f_0(1370)$ and $f_0(1500)$ contributions are included
in this fit (and the next).
The scattering length $a_0$ and the effective range  are
fitted with the errors quoted by CCL.
Parameters are shown in column (ii) of Table 1.

Resulting errors decrease compared with
the BES publication  because including $\sigma \to KK$ and $\eta
\eta$ produces cusps at these thresholds and removes the
requirement for the $f_0(980)$ contribution completely; the
errors of $\pm 30$ MeV in real and imaginary parts of the pole position
are taken from a range of fits to BES data using a variety of
fitting functions going beyond those used here; they include systematic
errors in data.
The shift in the pole arises partly from the use of
$j_1(s)$ of eqn. (6) but mostly from including $KK$ and $\eta \eta$
channels.
There is a distinct improvement in the fit to BES data.
This fit is used in making Fig. 1.
It is a valuable feature of the BES data that the $f_0(980) $
contribution is negligible, giving an unimpeded view of the
$\sigma $ pole.

The fit to phases is shown by the chain curve of Fig. 3
and the fit to BES data by the dashed curve of Fig. 4.
The fitted value of $a_0$ falls to $0.189 m_\pi ^{-1}$,
i.e. $6.2\sigma$ below the ChPT prediction.
The question is whether to believe the ChPT prediction or the
experimentally fitted scattering length.
The ChPT prediction was made using the Roy equation which
gives the low mass of CCL's $\sigma $ pole.
It is obvious that there will be a correlation between
$a_0$ and the pole position: as the pole moves away from
the $\pi \pi$ threshold, the scattering length will naturally
decrease.
If the scattering length is forced upwards, one finds an almost
linear relation between the pole mass and the scattering length.
To reach a value $a_0 = 0.215m_\pi ^{-1}$ requires a pole
mass of 472 MeV.
The $\chi^2$ of the fit increases substantially, mostly because
of the $K_{e4}$ and Cern-Munich data; in particular, the
382 MeV point of $K_{e4}$ data pulls the scattering length down.
The BES data at low mass also favour a small scattering length,
but are in a mass range where the available phase space is cutting
off the signal.

It must be said that this fit is not using information from the
Roy equation.
It is made empirically to data above threshold.
In the next step, the information from the Roy
equation is incorporated as closely as possible.

\subsection {A compromise solution}
In order to constrain the fit to data as closely as
possible to the Roy equation, a combined
fit is made to the $T_{11}$ of CCL up to 1.15 GeV, and data from
BES, $K_{e4}$ and Cern-Munich.
Errors are assigned to CCL phases rising linearly with
lab kinetic energy from zero at threshold to $3^\circ$ at 750
MeV.
This gives maximum weight to results from the Roy equation
near threshold, but allows flexibility in the effects of
$f_0(980)$, and continuations of effects of $KK$ and $\eta \eta $
inelasticities into the mass range 600 to 1000 MeV.
The pole moves to (472 - i 271) MeV and parameters are given in
column (iii) of Table 1.
The scattering length is $0.215 m_\pi ^{-1}$.
The fit to BES data is shown by the full curve of Fig. 4;
it is marginally poorer in terms of total $\chi^2$ but is obviously
acceptable.
However, the fit resists strongly any attempt to move the mass of the
pole any lower and the scattering length correspondingly higher.

The fit to phases is shown by the dashed curve of Fig. 3.
Up to 750 MeV, the difference between this fit and CCL phases is only
$1.2^\circ$ at 470 MeV.
This illustrates how difficult it is to deduce
the $\sigma$ pole position from elastic data.
The difference of  $1.65^\circ$ at 865 MeV is well within errors of
data.

Strictly speaking, the small perturbation to the $\sigma$ amplitude on
the right-hand cut induces, via crossing, a small perturbation to the
left-hand cut.
However, the $\sigma$ contribution on the left-hand cut has a small
isospin coefficient, and I have checked that in practice the
perturbation is smaller than errors in the large contributions from
$\rho$ and $f_2$ exchange.

What scattering length should be adopted?
The lowest order prediction from Weinberg's current algebra [20]
is $0.16m_\pi^{-1}$.
In second order ChPT it rises to $0.20m_\pi^{-1}$ and then
$0.215m_\pi^{-1}$ at fourth order.
The prediction of $(0.220 \pm 0.005)m_\pi^{-1}$ at sixth order by
Colangelo et al. [17] takes account
of the unitarity branch cut at threshold; however, it does depend
on using the Roy equations.
It seems safe to constrain the scattering length to be at
least $0.215m_\pi^{-1}$, so the conclusion is that
fit (iii) is currently the best compromise.
The change to the pole mass predicted by CCL is modest, but twice
the error they quote.

The full and dashed curves of Fig. 4 are very similar.
This is because they differ only in the scattering length $a_0$.
An interesting result is that, for all three fits, the effective
range does not change significantly from the CCL value.
In lowest order ChPT, the effective range is proportional to
\begin {equation}
2a_0 - 5a_2 = 3m_\pi^2/4\pi F^2_\pi,
\end {equation}
as shown by Weinberg [19], so this relation is accurately
consistent
with all three fits. In this relation, $F_\pi$ is the pion decay
constant.

\section {Possible ambiguities in fitting BES data}
Although the effect of systematic changes in inelasticity over
the mass range from threshold to 1.15 GeV provides
a natural resolution of the discrepancy between BES data and the CCL
prediction, alternatives have been suggested.
Two points have appeared in recent preprints.
Firstly, Wu and Zou [21]
remark that the width fitted to $b_1(1235)$ is 195 MeV compared with
the value $142 \pm 9$ MeV of the Particle Data Group [22].
This discrepancy was also reported by DM2 [2].
Wu and Zou suggest that the strong process $J/\Psi \to \rho \pi$,
followed by $\rho \to \omega \pi$ may contribute.
In the original BES work, the possibility $\rho (770)
\to \omega \pi$ was tried and gave little improvement and, at maximum,
a contribution of 2\% (intensity) of the data.
In any case, the effect lies close to  the vertical and horizontal
$b_1$ bands of Fig. 1(a), particularly the region where they cross
near the bottom left-hand corner of the Dalitz plot.
This is remote from the $\sigma$ band; changes to interferences
between $\sigma$ and $b_1$, $\rho$ or $\rho '$
have negligible effect on $\sigma $ parameters.

Secondly, Caprini [23] suggests that triangle graphs due to
$b_1 \to \omega \pi$, followed by $\pi \pi$ rescattering will
introduce effects beyond the isobar model.
Although this is true, it is known that such effects vary
logarithmically over the Dalitz plot.
There is no obvious reason why they should introduce a rapidly
varying effect close to the right-hand edge of the Dalitz plot.

Thirdly, could there be some form of background in $J/\Psi \to
\omega \sigma$?
If such a background is included as a quadratic
function of $s$, the discrepancy with CCL persists.
The reason is that BES  data determine the $\pi \pi$ S-wave amplitude
accurately above the  $f_2(1270)$ as well as below it and limit it to
small values; this  severely limits the form of any background to
something close to the observed peak.
In particular, a conventional form factor $\exp (-k^2R^2/6)$ at the
vertex for $J/\Psi \to \omega \sigma$ (where $k$ is
$\omega$ momentum in the $J/\Psi$ rest frame) gives no improvement,
since it requires negative $R^2$. So a background does not provide a
simple escape route.

A fourth possibility has been raised in discussions with CCL.
This is that the form factor has a zero somewhere above 1 GeV.
Extrapolating the full curve of Fig. 4, there might in principle
be a zero in the mass range around 1.3 GeV.
This would be obscured in the BES data on $J/\Psi \to \omega \pi \pi$
by the $f_2(1270)$.
However, there are also data on $J/\Psi \to \omega K^+K^-$ where the
$\sigma \to KK$ amplitude is clearly required [24].
The $ f_2(1270) \to KK$ contribution in those data is sufficiently
small that a zero in the $\sigma \to KK$ amplitude can be ruled
out definitively up to $\sim 1.6 $ GeV.
At that mass, the required radius of interaction would be
unreasonably large, $> 0.8$ fm, and the $\sigma$ pole region
would be seriously distorted by the form factor.
Data on production of the $\kappa$ pole in $J/\Psi \to
K^*(890)K\pi$ require an RMS radius $<0.38$ fm with 95\% confidence
[25].

\section {Discussion and conclusions}
There is a significant conflict in Fig. 4 between the
$\sigma$ pole of CCL and BES data.
In my view, the discrepancy needs explanation and the
BES data should be taken at face value.
The prescription adopted here in eqns. (4), (8) and (9)
gives a natural improvement in fitting BES data without
disturbing elastic phases up to 750 MeV by more than $1.2^\circ$
in fit  (iii).
This illustrates the ease with which the prediction of CCL may be
modified to fit BES data.

    The strength of the BES data is that a peak is clearly visible.
There is no significant $f_0(980)$ signal.
The BES data are therefore free of uncertainties about mixing
between $\sigma$ and $f_0(980)$.
The weakness of elastic scattering
data is that there is no visible peak which can be checked, and there
is a significant $f_0(980)$ amplitude which must be separated.
The BES data provide a better view of the upper side of the $\sigma$
pole than elastic data, where the low mass tail of $f_0(980)$ is
uncertain.
The CCL calculation provides a better view of its lower side, where
constraints from ChPT are valuable.

A more ambitious approach, beyond the scope of present work, would
be a solution of coupled channel Roy equations for
$\pi \pi$, $KK$ and $\eta \eta$ channels, including the dynamics
driving $f_0(980)$ and $\sigma \to KK$ and $\eta \eta$.
However, from experience with the model of van Beveren et al
[19], it is likely that uncertainties are presently too large
to give a definitive prediction of the dynamics of $f_0(980)$
and its delicate mixing with the $\sigma$.

\vskip 1.2mm

I am grateful to Prof. H. Leutwyler for extensive discussions of
the details of the calculation of Ref. [1].
I am also indebted to Profs. G. Rupp and E. van Beveren
for use of their programme and Dr. F. Kleefeld for illuminating
discussions.

\end {document}